\documentclass[prb,preprint]{revtex4}
\usepackage[utf8]{inputenc}
\usepackage[english]{babel}
\usepackage{amsmath}
\usepackage{amsfonts}
\usepackage{amssymb}
\usepackage{indentfirst}
\usepackage{textcomp}
\usepackage{epsfig}
\usepackage{subfigure}
\usepackage{listings}
\usepackage{multirow}
\usepackage{graphicx}
\usepackage{cleveref}
\newcommand{\superscript}[1]{\ensuremath{^{\textrm{#1}}}}

\usepackage{array,booktabs}
\newcolumntype{C}[1]{>{\centering\arraybackslash}p{#1}}
\usepackage{chemformula} 
\begin{document}
\title{Plasmon-Polaritons in Nanoparticle Supercrystals: Microscopic Quantum Theory Beyond the Dipole Approximation}
\author{Eduardo B. Barros\superscript{1}}
\author{Bruno Gondim Vieira\superscript{1}}
\author{Niclas S. Mueller \superscript{2}}
\author{Stephanie Reich \superscript{2}}

\affiliation{\superscript{1} Department of Physics,Universidade Federal do Cear\'a, Fortaleza, Cear\'a, 60455-760 Brazil}

\affiliation{\superscript{2} Department of Physics, Freie Universität Berlin, Arnimallee 14, D-14195 Berlin, Germany}

\date{\today}

\begin{abstract}
Crystals of plasmonic metal nanoparticles have intriguing optical properties. They reach the regimes of ultrastrong and deep strong light-matter coupling, where the photonic states need to be included in the simulation of material properties. 
We propose a quantum description of the plasmon polaritons in  supercrystals that starts from the dipole and quadrupole excitations of the nanoparticle building blocks and their coupling to photons. Our model excellently reproduces results of finite difference time domain simulations. It provides detailed insight into the emergence of the polariton states. Using the example of a face centered cubic crystals we show that the dipole and quadrupole states mix in many high symmetry directions of the Brilouin zone. A proper description of the plasmon and plasmon-polariton band structure is only possible when including the quadrupole-derived states. Our model leads to an expression of the reduced coupling strength in nanoparticle supercrystals that we show to enter the  deep strong coupling regime for metal fill fractions above $0.8$. In addition to the plasmon-polariton energies we analyse the relative contributions of the dipole, quadrupole, and photonic states to their eigenfunctions and are able to demonstrate the decoupling of light in the deep strong coupling regime. 
Our results pave the way for a better understanding of the quantum properties of metallic nanoparticle supercrystals in the ultrastrong and deep-strong coupling regime. 
\end{abstract}

\maketitle

\section{Introduction}
\label{sec:intro}

Nanoparticle supercrystals are three-dimensional  lattices of nanoparticles with long-range crystalline order.\cite{Murray1980,boles_self-assembly_2016,murray_synthesis_2000,Shevchenko2006,Coropceanu2019,Mueller2020,Schulz2020} They have long fascinated physicists, chemists and material scientists, because they promise material properties that cannot be found in nature.\cite{boles_self-assembly_2016} Supercrystals are tailored through particle-particle interaction and the choice of their building blocks. Initially, the discussion  focused on collective vibrational and electronic modes of supercrystals made from semiconducting nanoparticles,\cite{Vovk2020} but interest recently turned to collective optical excitations,\cite{Baimuratov2013} especially, when using metallic nanoparticles as the  supercrystal building blocks.\cite{Mueller2020,Schulz2020, Mueller2021,GarciaLojo2019,BlancoFormoso2020} The optical properties of metallic nanoparticle supercrystals are dominated by the response of their free electrons. 
When light interacts with a metallic nanoparticle, it excites localized surface plasmons - the collective oscillation of the free electrons - that strongly absorb and scatter light.\cite{Kreibig,Kelly2003, MaierBook} Plasmons of different nanoparticles interact in supercrystals over various length scales creating coherent collective excitations that propagate through the lattice.\cite{Meinzer2014,Lamowski2018,Mueller2020} The collective plasmonic modes couple to photons forming hybrid quasi-particles called plasmon polaritons.\cite{Barnes2003} Supercrystal plasmon polaritons differ greatly in their properties from the excitations of the individual nanoparticles, they determine the optical response of metallic supercrystals, and open new pathways for scientific discoveries and technological development.\cite{Shalaev2008,Tame2013,Vovk2020,Mueller2020} 

Recently, we showed that plasmonic supercrystals can be used to explore phenomena in the ultrastrong regime  of light-matter coupling (USC).\cite{Mueller2020,Mueller2021,Baranov2020} In this regime the coupling strength is a considerable fraction of the bare frequency of the system, which leads to peculiar properties of the polariton states.\cite{Kockum2019,FornDiaz2019} For inter-particle gaps much smaller than the nanoparticle size, light-matter coupling even enters the regime of deep strong coupling (DSC),\cite{Mueller2020} where the interaction between light and matter exceeds the energies of the bare excitation. This means that the properties of plasmonic nanoparticle supercrystals can only be modeled when considering the existence of photonic states.\cite{Mueller2020} This is in sharp contrast to the standard treatment of light as an external perturbation. 
The DSC regime, moreover, promotes a wide range of exquisite and interesting physical effects such as a decoupling of light and matter and the breakdown of the Purcell effect \cite{DeLiberato2014,Mueller2020}, the squeezing of the photonic components of the polaritons, super-Poissonian phonon and photon statistics \cite{Artoni1991}, and ground state electroluminescence.\cite{Cirio2016,Kockum2019,FornDiaz2019} Although most of these properties can be understood in terms of the Hopfield model for light-matter interaction \cite{Hopfield1958}, a microscopic model that is capable of making predictions for a specific plasmonic supercrystals is highly desirable.

Weick {\it et al.}\cite{Weick2015} and Lamowski {\it et al.}\cite{Lamowski2018} developed a quantum-based formalism that describes the polaritons of plasmonic supercrystals. The microscopic structure of the supercrystal and the dipole-dipole interactions between the nanoparticles turned out to be key for modelling the collective plasmon modes.\cite{Lamowski2018} The model uses two common approximations that appear very reasonable on first sight, but turn out to limit its applicability: It considers only the dipole excitation of plasmonic nanoparticles and neglects Umklapp processes when crossing the boundary of the Brillouin zone. The restriction to the dipole excitation is motivated by the small size ($<100\,$nm) of plasmonic nanoparticles that prohibits the excitation of higher-order electrical modes in individual particles.\cite{MaierBook,LeRuBook} This argument, while correct, misses that higher-order modes of individual nanoparticles combine into dipole-active eigenstates in plasmonic oligomers and supercrystals.\cite{Reich2020,LiJensen2006} These states may couple to the dipole-induced collective plasmons and the electromagnetic states affecting the final polariton dispersion. Umklapp processes are usually negligible for optical excitations, because the wavelength of visible light ($500\,$nm) is small compared to the translational periodicity in natural crystals ($0.1\,$nm). The unit cell of supercrystals, however, becomes a sizable fraction of the light wavelength, since nanoparticle diameters are several $10\,$ to $100\,$nm. The quasi-static approximation breaks down and Umklapp processes may turn out to be important. 

In this work we develop a microscopic quantum model for plasmon-polaritons in metallic supercrystals that includes quadrupolar plasmonic excitations and Umklapp processes. We validate it by comparison to finite difference time domain (FDTD) simulations. Our model very well describes plasmon polaritons for supercrystals with low and high packing densities. We calculate the plasmon band structure for face centered cubic (FCC) supercrystals as a function of packing density showing that the quadrupole-derived collective eigenmodes cross and mix with the dipole-induced states for high packing densities. Including the coupling to the electromagnetic states results in plasmon polaritons in the USC and DSC regimes. The strongest contribution to light-matter coupling arise from the dipole plasmons; the quadrupole-photon coupling is an order of magnitude weaker than dipole-photon interaction. However, the quadrupole contribution is important for the polariton band structure, because the energies of the collective plasmons are overestimated by several 100\,meV at the Brillouin zone boundary when neglecting the quadrupole modes. We derive a closed expresssion for the reduced coupling strength and show that is mainly depends on the metal fill fraction. We extract the dipole, quadrupole, and photon contribution to all polariton states. The decoupling of light and matter clearly manifests as the three quasi-particles dominate distinct polariton branches in the DSC regime.

This paper is organized as follows: In Sec.~\ref{sec_QPPM}, we describe the theoretical framework and apply it to a three-dimensional Bravais lattice of spherical metallic nanoparticles. The theory for arbitrary crystal structures is given in the Supplementary Information. We compare the calculated polariton dispersion to FDTD simulations. In Sec.~\ref{sec:results} we calculate the plasmon and plasmon-polariton band structure of FCC supercrystals. We demonstrate how quadrupole modes and light-matter coupling affect the polariton dispersion. We discuss  the properties of polaritons, the reduced coupling strength, and demonstrate the decoupling of light and plasmons in the DSC regime. In Sec.~\ref{sec:conclusions} we summarize the main findings of the paper. 

\section{Quantum microscopic plasmon-polariton model\label{sec_QPPM}}

In this section we derive the microscopic model of plasmon polaritons in metallic supercrystals. We first present a general theoretical framework that describes individual and interacting nanoparticles with dipole and quadrupole excitations and their coupling to an electromagnetic field. This description is then applied to a Bravais lattic; we verify its validity and limitations by comparing to FDTD simulations of FCC nanoparticle crystals. Our microscopic quantum plasmon-polariton model contains the nanoparticle quadrupole in addition to their dipole excitations. Adding the quadrupole terms was challenging, because the the quantum description of light-matter interaction is based on the dipole approximation. 
One difficulty we encountered was to find a proper description of the conjugate momenta for higher-order multipoles. This problem dates back to the description of nuclear excitations and was discussed first by Bohr and Mottelson.\cite{Bohr1953,Bohr1998} In 1978 Gulshani finally provided a formal development of canonically conjugate momenta for quadrupolar excitations,\cite{Gulshani1978} but no  solution has been found for higher-order multipoles. 

We start from the most general Hamiltonian for a set of charges distributed in space interacting with the electromagnetic field
\begin{equation}\label{eq_fullH_LM}
\mathcal H=\sum_n \frac{1}{2m}\left[\vec p_n - {\rm q}_n \vec A(\vec r_n, t)\right]^2 + V_{Coul} + \mathcal{H}_L,  
\end{equation}
where $V_{Coul}$ is the Coulomb interaction between the different $n$ charges ${\rm q}_n$ in the system, $\mathcal{H}_L$ is the quantized free-electromagnetic field Hamiltonian and $\vec p_n$ is the conjugate momentum to $\vec r_n$. $\vec A$ is assumed to be in the Coulomb gauge.   
We consider an isolated spherical metallic nanoparticle at the origin. The charges are bound by the nanoparticle volume, such that the summation $n$ in Eq.\ (\ref{eq_fullH_LM}) is restricted to the free electrons in the particle. We  define a set of variables
\begin{equation}
h_{\sigma} =\frac{1}{N} \sum_{n} r_{n,\sigma},\:H_{\gamma} =\frac{1}{N\bar \rho} \sum_{n,\alpha,\beta}r'_{n,\alpha} r'_{n,\beta}\chi^\gamma_{\alpha\beta},  
\end{equation}
where $h_{\sigma}$ represents the center of mass displacement along the $\sigma$ direction and is associated with the dipole moment of the charge distribution. $n=1,...,N_j$ runs through all the charges in a given nanoparticle. The second term $H_{\gamma}$ is associated to the quadrupole moment, with $\alpha$ and $\beta$ being different Cartesian directions. $\gamma$ specifies one of the five possible quadrupolar modes, and $r'_{n\alpha}=r_{n,\alpha}-h_{\alpha}$. $\bar \rho=\langle 1/N\sum_{n}r_{n}^2\rangle \sim \sqrt{\frac{2}{5}}\rho$, 
with $\rho$ the radius of the spherical particle, is the expectation value of the diagonal term of the quadrupole moment. The respective conjugate momenta are
\begin{equation}
\pi_{\sigma}=\sum_{n} p_{n,\sigma} ,\: \Pi_{\gamma}= \frac{1}{\bar \rho}\sum_{n,\alpha,\beta} p'_{n,\alpha} r'_{n,\beta}\chi^\gamma_{\alpha\beta},
\end{equation}
with $p'_{n,\alpha}=p_{n,\alpha}-\pi_{\alpha}$. It should be mentioned that only the dipole part of this transformation is formally canonical. As discussed in the S.I., the quadrupole term can only be associated with a canonical transformation (and thus with the expected commutation relations) if the total angular momentum of the charge distribution is zero and if the charge displacements are small compared to the bulk charge.\cite{Gulshani1978} With these two considerations, the dynamical variable $\sum_n r_{n,\alpha}^2$ can be substituted by its expectation value $\bar \rho$ and the quadrupolar moments can be described in terms of the traceless and symmetric $3\times3$ matrices $\chi_\gamma$ which act as unit tensors within the dyadic double-dot product (see S.I.). With this approximation, we can write the following inverse transformation for the quadrupolar canonical variables
\begin{equation}
\left\{\begin{matrix}
\sum_{n,\alpha,\beta}r_{n,\alpha}r_{n,\beta}=N\bar\rho^2\left({\bf 1} + \frac{1}{\bar\rho}\sum_\gamma  H_{\gamma}\chi_\gamma\right)\\
\sum_{n,\alpha,\beta}p_{n,\alpha}r_{n,\beta}=\bar\rho \sum_\gamma \Pi_{\gamma}\chi_\gamma
\end{matrix}\right. .
\end{equation}
Here, ${\bf 1}$ is the $3\times 3$ unit matrix.    

We now consider a system of many nanoparticles. If there is no charge exchange between different particles, the full Hamiltonian of the system will be given by Eq.\ (\ref{eq_fullH_LM}) but with the summation in $n$ being extended to a summation for each particle $n\in j$ and a summation of the different particles in the system. With this, the matter Hamiltonian ($\mathcal{H}_M=\sum_n \frac{p^2_n}{2m} + V_{Coul}$) can be written as
\begin{equation}
\mathcal{H}_M=\mathcal{H}_M^D+\mathcal{H}_M^Q+\mathcal{H}_{\rm plpl}+\mathcal{H}^{HO}_M, 
\end{equation}
where $\mathcal{H}_M^D + \mathcal{H}_M^Q$ represents the Hamiltonians for each nanoparticle in the system, including the kinetic energy associated with each of the canonical coordinates ($D$ for dipole and $Q$ for quadrupole) and the intra-particle Coulomb interactions. Both are assumed to be described in terms of harmonic oscillations with characteristic frequencies $\omega_{j,D}$ and $\omega_{j,Q}$,  which can have different values for each nanoparticle $j$ in the system. The term $\mathcal{H}_M^{HO}$ represents the dynamics of the higher-order coordinates, which are disregarded in the present model.
 
The plasmon-plasmon interaction between the different particles in the system is
\begin{equation}
\mathcal{H}_{\rm plpl}=\mathcal{H}_{DD}+\mathcal{H}_{DQ}+\mathcal{H}_{QQ},
\end{equation}
corresponding to the dipole-dipole $\mathcal{H}_{DD}$, dipole-quadrupole $\mathcal{H}_{DQ}$ and quadrupole-quadrupole interaction $\mathcal{H}_{QQ}$. Explicit expressions for these terms are given in the Supplementary Information.
  
The characteristic size of plasmonic nanoparticles is $10-100\,$nm. The particles diameters are a considerable fraction of the light wavelength. The common approach for  light-matter interaction of expanding the vector-potential in a Taylor series in the vicinity of the charge distribution and disregarding higher-order terms in $k$ will not be effective. We perform the expansion in a slightly different way; we start from the vector potential of a plane wave given by
\begin{equation}
A_\lambda(\vec r,t)=\sum_{\vec q}A_{\vec q,\lambda}(t)\exp(i\vec q\cdot\vec r).
\end{equation}
where $\lambda$ specifies the light polarization and $\vec q$ is a vector in reciprocal space - not to be confused with the charges $\rm q$. With this, the general light-matter interaction for a particle $j$ can be separated into two parts. The first-order part 
\begin{equation}
\mathcal{H}_{LM}^{(1)}=-\frac{\rm q_e}{m_e}\sum_{j,(n\in j),\lambda,\vec q}p_{n\lambda} A_{\vec q\lambda}\exp(i\vec q\cdot\vec r_n),
\end{equation}
corresponds to the interaction of light with excitations of the electric charge distribution. The time dependence of $\vec A$ is implicit. The second-order part 
\begin{equation}
\mathcal{H}_{LM}^{(2)}=\frac{\rm q^2_e}{2m_e}\sum_{j,n \in j,\lambda,\vec q,\vec q'}A^{\ast}_{\vec q'\lambda} A_{\vec q\lambda}\exp[i(\vec q-\vec q')\cdot\vec r_n],
\end{equation}
describes the back reaction of the electric field as it accelerates the electric charges.\cite{Selsto2007} This term
is known as the $A^2$ term in the light-matter Hamiltonian; it
becomes extremely important in the USC and DSC coupling regime.\cite{Kockum2019, Mueller2020} 

Let us now take the spherical harmonic expansion of the plane wave around the position $\vec R_j$ of each nanoparticle in the system
\begin{equation}
\exp(i\vec q\cdot\vec r_j)\sim j_0(qr_j) + 3i \frac{j_1(qr_j)}{qr_j}\vec q\cdot \vec r_j,
\end{equation}
where $\vec r_j=\vec r-\vec R_j$ and we have retained only the lower order terms. We apply this expansion to the 1st order part of the light-matter interaction for each nanoparticle $j$ independently and sum to get the full Hamiltonian. It then has a dipole-like contribution
\begin{equation}
\mathcal{H}_{LM}^{(D,1)}=-\frac{\rm q_e}{m_e}\sum_{j,\lambda,\vec q}A_{\vec q\lambda}\exp(i\vec q\cdot \vec R_j)\sum_{n\in j}p_{n\lambda} j_0(qr_{n,j}), 
\end{equation}
and a quadrupolar-like term  
\begin{equation}
\begin{split}
\mathcal{H}_{LM}^{(Q,1)}=&
-\frac{\rm q_e}{m_e}\sum_{j,\lambda,\vec q,\alpha,\beta}  iqA_{\vec q\lambda}\exp(i\vec q\cdot \vec R_j)\times \\&\times\left(\hat e_\lambda \hat e_q:\hat e_\alpha \hat e_\beta\right) \sum_{n\in j}\frac{3j_1(qr_n)}{qr_n} p_{n\alpha}r_{n\beta},  
\end{split}
\end{equation}
where $(:)$ stands for a double-dot dyadic product. 
We now substitute $j_0(qr_n)$ and $3j_1(qr_n)/qr_n$ by their mean values in each nanoparticle 
\begin{equation}
{\rm f}_D=\langle j_0(qr_j) \rangle =\frac{3}{(q\rho_j)^3}\left[\sin(q\rho_j)-q\rho_j \cos(q\rho_j)\right]
\end{equation}
and
\begin{equation}
{\rm f}_Q=\langle \frac{3j_1(qr_n)}{qr_n}\rangle=\frac{9}{(q\rho_j)^3}\left[{\rm Si}(q\rho_j)-\sin(q\rho_j)\right],   
\end{equation}
where Si$(x)$ is the sine integral function. ${\rm f}_D$ and ${\rm f}_Q$ are nanoparticle form factors somewhat similar to the atomic form factors in X-ray diffraction theory\cite{Kittel2005}.
The magnitude of the form factors decrease with increasing $q\rho$, effectively cutting off the contribution of photons with wavevectors much larger than $1/\rho$ to the light-matter interaction. This effect stems from the field retardation within the nanoparticle.

With this definition the 1st-order term in the light-matter interaction Hamiltonian becomes
\begin{equation}
\mathcal{H}_{LM}^{(D,1)}=-\sum_{j,\vec q}{\rm f}_D(q)\frac{{\rm Q}_{D}}{M}\vec \pi_j\cdot \vec A_{\vec q,\vec R_j},
\end{equation}
and
\begin{equation}
\mathcal{H}_{LM}^{(Q,1)}=\sum_{j,\vec q}{\rm f}_Q(q)\frac{ {\rm Q}_{Q}\bar \rho_j}{M} \Pi_{j,\nu}[\chi_\nu : \vec q\vec A_{\vec q,\vec R_j}], 
\end{equation}
where $M$ is the total mass of the charges ($M=Nm_e$), ${\rm Q}_D$ and ${\rm Q}_Q$ are the screened dipole and quadrupole effective charges, which will depend on the relative permittivity of the surrounding medium, $\vec q \vec A_{\vec q,\vec R_j}$ is a dyadic, with $A_{\vec q,\vec R_j}=A_{\vec q}\exp(i\vec q\cdot\vec R_j)$. 
Note that for small nanoparticle radii ($\rho$) the form factors approach unity and the plasmon-photon interaction obtained by applying Taylor's expansion is recovered. In that case, 1st order light-quadrupole interaction increases linearly with $q$. For the second-order term, associated to the $A^2$ light-matter interaction term, we have the two contributions
\begin{equation}
\mathcal{H}_{LM}^{(D,2)}=\sum_{j,\vec q,\vec q'}{\rm f}_D(q){\rm f}_D(q')\frac{{\rm Q}^2_{D}}{2M}\left[\vec A^\ast_{\vec q',\vec R_j}\cdot \vec A_{\vec q,\vec R_j}\right],
\end{equation}
\begin{equation}
\mathcal{H}_{LM}^{(Q,2)}=\sum_{j,\vec q,\vec q'} {\rm f}_Q(q){\rm f}_Q(q')\frac{{\rm Q}^2_{Q}\bar\rho_j^2}{2M}\left[\vec A^\ast_{\vec q',\vec R_j}\vec q'\cdot \vec q\vec A_{\vec q,\vec R_j}\right].
\end{equation}
The terms involving products between $\vec A_{\vec q}$ and $\bar\nabla \vec A_{\vec q}$ are disregarded, as the expectation value $\langle \sum_n\vec r_n\rangle=0$ vanishes. Also, all non-quadratic terms, involving more than two dynamical variables, are disregarded within this approximation.    
  
\subsection{Plasmonic nanoparticle crystals}\label{sec:NPcrystals}

We now apply the proposed quantum mechanical description of light-matter coupling with dipole and quadrupole modes to a crystal of identical spherical nanoparticles placed in a Bravais lattice. The position of a nanoparticle in the crystal is determined by a lattice vector $\vec R$, and the index $j$ is dropped. The model can be trivially extended to crystals with an arbitrary basis - see Supplementary Information for equations. 

Following the work of Weick~{\it et al.}\cite{Weick2015} and Lamowski~{\it et al.}\cite{Lamowski2018}, we expand the plasmonic and photonic dynamical variables into creation and annihilation operators defined in reciprocal space for a periodic arrangement of particles. We obtain the Hamiltonian for plasmon-plasmon interaction as
\begin{equation}
\begin{split}
\mathcal{H}_{\rm plpl}=&\sum_{\vec q,\nu,\nu'}\hbar\sqrt{\Lambda_{\bar\nu}\Lambda_{\bar\nu'}}S_{\nu,\nu'}^{\bar\nu,\bar\nu'}(\vec q)\left(b_{-\vec q,\nu}^\dag+b_{\vec q,\nu}\right)\times\\&\times\left(b_{\vec q',\nu'}^\dag+b_{-\vec q',\nu'}\right),
\end{split}
\label{eq_Hplpl}
\end{equation}
where $b_{\vec R,\nu}=1/\sqrt{N_\mathrm{cells}}\sum_{\vec q} b_{\vec q,\nu}\exp(i\vec q\cdot \vec R)$ is the annihilation operator for multipole  oscillations of the nanoparticle in the unit cell defined by the lattice vector $\vec R$. $\bar\nu=D$ for $\nu=1-3$ which correspond to dipole modes while $\bar\nu=Q$ for the $\nu=4-8$ quadrupole modes.
$N_\mathrm{cells}$ is the number of unit cells. 
The structure function $S_{\nu,\nu'}^{\bar\nu,\bar\nu'}$ in Eq.~\eqref{eq_Hplpl} depend only on the Bravais lattice; it is given by
\begin{widetext}
\begin{equation}
\label{eq_FDD}
S^{DD}_{\nu,\nu'}(\vec q)=\sum_{\vec{R}} \frac{1}{2}\frac{\delta_{\nu\nu'} -3 (\hat e_\nu\cdot\vec{n})(\hat e_{\nu'}\cdot\vec{n})}{(R/\bar R)^3}\exp(i\vec{q}\cdot\vec R), 
\end{equation}
when both $\nu$ and $\nu'$ correspond to dipole modes,
\begin{equation}
\label{eq_FQQ}
S^{QQ}_{\nu,\nu'}(\vec q)=\sum_{\vec{R}}\frac{1}{6} \left\{35\frac{(\chi_\nu:\hat n \hat n)(\chi_{\nu'}:\hat n\hat n)}{(R/\bar R)^5} - 20\frac{(\chi_\nu\chi_{\nu'}:\hat n\hat n)}{(R/\bar R)^5}+2\frac{(\chi_\nu:\chi_{\nu'})}{(R/\bar R)^5}\right\} \exp(i\vec{q}\cdot\vec R),
\end{equation} 
when both $\nu$ and $\nu'$ correspond to quadrupole modes, and
\begin{equation}
\label{eq_FDQ}
S^{DQ}_{\nu,\nu'}(\vec q)=\sum_{\vec{R}}\frac{1}{2}\left[  
-5\left(\chi_{\nu'} : \frac{\hat n \hat n}{(R/\bar R)^4}\right)(\hat n \cdot \hat e_\nu) + 2 \left(\chi_{\nu'} : \frac{\hat n \hat e_\nu}{(R/\bar R)^4}\right)\right] \exp(i\vec{q}\cdot\vec R),
\end{equation} 
\end{widetext}
when $\nu$ corresponds to a dipole mode and $\nu'$ corresponds to a quadrupole mode. Here $\hat a\hat b$ corresponds to dyadics formed by the two unit vectors $\hat a$ and $\hat b$. Also, $\hat n=\vec R/R$ and $\bar R = (V_{uc})^{1/3}$, with $V_{uc}$ being the volume of the unit cell. The dipole-dipole interaction does not converge  for $q\rightarrow 0$ in a filled three-dimensional space.\cite{Lamowski2018,Cohen1955} For wavevectors below a cutoff value $q_c$, {\it i.e.}, $|q|<|q_c|$, the dipole-dipole structure function $S^{DD}$ is replaced by
$S^{DD}_{\nu,\nu'}=-2\pi\left[\delta_{\nu\nu'}-(\hat e_{\nu}\cdot \hat q)(\hat e_{\nu'}\cdot \hat q)\right]/3$. The value of $q_c$ that allows for a smooth dispersion relation depends on the Bravais lattice and on the number of unit cells considered. The plasmon-plasmon interaction Hamiltonian in Eq.~\eqref{eq_Hplpl} contains coupling factors $\Lambda_{\bar\nu}$; they are given by
\begin{equation}
\Lambda_{D}=\frac{ {\rm Q}_{D}^2}{8\pi\epsilon_0\epsilon_m M\omega_D V_{uc}},
\end{equation} 
and 
\begin{equation}
\Lambda_{Q}=\frac{({\rm Q}_{Q}\bar\rho)^2}{8\pi\epsilon_0\epsilon_m M\omega_QV_{uc}^{5/3}},
\end{equation} 
where $\epsilon_m$ is the dielectric constant of the surrounding medium, which is assumed to be a positive constant. 

We now turn to the interaction between plasmons and photons. The first-order part of the plasmon-photon coupling can be written as  
\begin{equation}
\begin{split}
\mathcal{H}_{\rm plpt}^{(1)}=i\hbar\sum_{\vec q,\vec G,\lambda,\nu} &\omega_{\bar\nu}\xi^{\nu}_{\lambda,\vec G}\left(b_{-\vec q,\nu}^\dag -b_{\vec{q},\nu} \right)\times\\&\times\left(c_{-\vec{q}-\vec G,\lambda}+c_{\vec{q}+\vec G,\lambda}^\dag\right),
\end{split}
\end{equation}
where $\vec G$ runs through the reciprocal lattice vectors for the chosen lattice. We defined
\begin{equation}
\label{eq_xiD}
\xi^{\nu}_{\lambda,\vec G}(\vec q)=\rm{f}_D(|\vec q+\vec G|)\xi_0^D(\vec q)P^D_{\nu,\lambda}(\vec q+\vec G),
\end{equation}
for $\bar\nu=D$ and 
\begin{equation}
\label{eq_xiQ}
\xi^{\nu}_{\lambda,\vec G}(\vec q)=i{\rm f}_Q(|\vec q+\vec G|)|\vec q+\vec G|\bar R\xi_0^Q(\vec q)P^Q_{\nu,\lambda}(\vec q+\vec G),
\end{equation}
for $\bar\nu=Q$, where
\begin{equation}
\xi_0^{\bar\nu}(\vec q)=\sqrt{\frac{2\pi\Lambda_{\bar\nu}}{\omega_{pt}(\vec q)}},
\end{equation}
and 
\begin{equation}
\left\{
\begin{array}{ll}
P^D_{\nu,\lambda}(\vec q) = &\hat e_\nu \cdot \hat e_\lambda \\
P^Q_{\nu,\lambda}(\vec q) = &\frac{1}{2}[\chi_\nu :\hat e_q \hat e_\lambda+\hat e_\lambda \hat e_q] 
\end{array}\right. ,
\end{equation}
with $\hat e_\lambda$ being functions of $\vec q$, since for both values of $\lambda$ the vector potential is perpendicular to the wavevector $\vec q$. 

Finally, the second-order part of the plasmon-photon interaction is
\begin{equation}
\begin{split}
\mathcal{H}_{\rm plpt}^{(2)}=\hbar\sum_{\vec q,\lambda, \lambda', \vec G,\vec G'} &\Xi^{\lambda\lambda'}_{\vec G,\vec G'}(\vec q)\left(c_{-\vec{q}-\vec G',\lambda'}^\dag+c_{\vec{q}+\vec G',\lambda'}\right)\times \\\times & \left(c_{-\vec{q}-\vec G,\lambda}+c_{\vec{q}+\vec G,\lambda}^\dag \right),
\end{split}
\end{equation}
where
\begin{equation}
\label{eq_Xi}
\Xi^{\lambda\lambda'}_{\vec G, \vec G'}(\vec q)=\sum_{\nu}\omega_{\bar\nu}\xi^{\nu\ast}_{\lambda'\vec G'}(\vec q)\xi^{\nu}_{\lambda\vec G}(\vec q).
\end{equation}
It is interesting to note that for each photon mode the second-order photon coupling is obtained in terms of a sum involving the matrix elements of the first-order interactions. In the weak-coupling regime, this term can be obtained using the TRK sum rule and has the important effect of balancing out the 1st-order term as $q$ goes to zero, thus preventing super-radiant phase transitions.\cite{Hepp1973}
Recently, a generalized sum rule was obtained for the strong coupling regime.\cite{Savasta2020} It is written in terms of the eigenstates of the full Hamiltonian and cannot be directly applied to simplify our calculations. However, the existence of such a rule indicates that even in the strong coupling regime, the second-order terms perfectly balance out the first-order interactions. It is also noteworthy that the second-order term directly couples photon modes with different polarizations $\lambda$ and $\lambda'$, effectively mixing these two otherwise independent photon states and opening pathways for different types of chiral activity in strongly coupled systems. 

To calculate the polaritonic modes, we can follow the work of Xiao\cite{Xiao2009}, and define a Bogoliubov vector operator 
\begin{equation}
\Phi_{\vec q}=
\begin{pmatrix} \bar b_{\vec q} \\ \bar c_{\vec{q}}\\ \bar b_{-\vec q}^\dag \\ \bar c_{-\vec{q}}^\dag  \end{pmatrix},
\end{equation}
where $\bar b_{\vec q}$, and $\bar b^\dag_{\vec q}$ are column vectors with each entry being an operator corresponding to a different plasmonic mode $\nu$. 
$\bar c_{\vec q}$, and $\bar c^\dag_{\vec q}$ are column vectors with operators for each polarization $\lambda$ 
and each reciprocal lattice vector $\vec G$ considered. The values of $\vec q$ are limited to the positive half of the Brillouin zone. $\Phi_{\vec q}$ obeys the following dynamical equation \cite{Xiao2009} 
\begin{equation}
i\hbar\frac{d}{dt}\Phi_{\vec q}=D_{\vec q}\Phi_{\vec q},
\end{equation}
with 
\begin{equation}
D_{\vec q}=\hbar\begin{pmatrix}
\bar \alpha_{\vec q} & \bar \gamma_{\vec q}\\
-\bar \gamma^\dag_{\vec q} & -\bar \alpha_{\vec q}^\star  
\end{pmatrix},
\end{equation}
where
\begin{equation}
\bar \alpha_{\vec q}=
\begin{pmatrix} \bar\omega_{pl}+\bar \Lambda (\bar S_{\vec q}+\bar S_{-\vec q}) & \bar\omega_{pl}\bar \xi_{\vec q}	\\
\bar\omega_{pl}\bar \xi^\dag_{\vec q} &\bar\omega^{pt}_{\vec q}+2 \bar \Xi_{\vec q}
\end{pmatrix},
\end{equation}
and
\begin{equation}
\bar \gamma_{\vec q}=
\begin{pmatrix} \bar\Lambda(\bar{S}_{\vec q}+\bar{S}_{-\vec q}) & \bar\omega_{pl}\bar \xi_{\vec q}(\vec q)	\\
-\bar\omega_{pl}\bar \xi^\dag_{\vec q} & 2 \bar \Xi_{\vec q}\end{pmatrix}.
\end{equation}
Here $\bar\omega_{pl}$ and $\bar\omega_{pt}$ are diagonal matrices with the energies of each of the plasmonic modes $\nu$ of the metallic nanoparticle and the photon modes (labelled by $\lambda$ and  $\vec G$)  that are taken into consideration. The matrix $\bar\Lambda$ is given by $\bar\Lambda=\sqrt{\Lambda_{\nu}\Lambda_{\nu'}}$, $\bar S_{\vec q}$ is given by Eqs.\ (\ref{eq_FDD})-(\ref{eq_FDQ}),  
$\bar \xi_{\vec q}$ by Eqs.\ (\ref{eq_xiD})-(\ref{eq_xiQ}), and $\bar \Xi_{\vec q}$ by Eq.\ (\ref{eq_Xi}). A detailed description of these matrices is given in the Supporting Information. 

The dynamical matrix $D_{\vec q}$ is diagonalized by a Bogoliubov-Valentin transformation $T_{\vec q}^{-1}D_{\vec q}T_{\vec q}$, which leads to a new set of creation and annihilation operators, $\Psi^\dag_{pp}(\vec q)$ and $\Psi_{pp}(\vec q)$, given by $\Psi_{pp}(\vec q)=T_{\vec q}\Phi_{\vec q}$  with eigenvalues $\hbar\omega_{pp}(\vec q)$.\cite{Xiao2009}
These operators correspond to the creation and annihilation of mixed excitations called plasmon-polaritons which have properties of both plasmons and photons.\cite{Lamowski2018} The eigenvalues can be associated with the plasmon-polariton dispersion. It should be mentioned that the transformation matrices $T_{\vec q}$ mix terms with both creation and annihilation operators of plasmons and photons, giving rise to many of the phenomena expected in the extreme regimes of light-matter coupling.\cite{Artoni1991,Cirio2016,Kockum2019,FornDiaz2019}  

\subsection{Quasi-static approximation}\label{sec:quasistatic}

In this section we discuss the input parameters of our microscopic model. We want to calculate the plasmon-polariton dispersions of plasmonic supercrystals and compare it to experimental results as well as calculations performed within other techniques. To do so, we need the frequencies of the dipole $\omega_D$ and quadrupole $\omega_Q$ plasmon resonances in metallic nanoparticles as well as their coupling factors $\Lambda_D$ and $\Lambda_Q$. 

The dipole and quadrupole frequencies are obtained within the quasi-static approximation. We consider a Drude metal with permittivity $\epsilon(\omega)=\epsilon_d-(\omega_p/\omega)^2$, neglecting losses. $\omega_p$ is the plasma frequency of the metal and $\epsilon_d$ a dielectric constant that accounts for the screening by bound charges.\cite{MaierBook} This yields the frequencies\cite{Kolwas2010,Shopa2010} 
\begin{equation}
\omega_D=\frac{\omega_p}{\sqrt{\epsilon_d+2\epsilon_m}},
\end{equation}
and
\begin{equation}
\omega_Q=\frac{\omega_p}{\sqrt{\epsilon_d+(3/2)\epsilon_m}}.
\end{equation}

The coupling parameters are obtained by considering the screened effective charges 
\begin{equation}
{\rm Q}_{l}=\frac{(2l+1)\epsilon_m}{l\epsilon(\omega)+(l+1)\epsilon_m} {\rm Q}_{l}^0,
\end{equation}
where $l=1$ for dipole and $l=2$ for quadrupole modes.\cite{Doerr2017} This leads to 
\begin{equation}\label{eq_lambd}
\Lambda_D=\frac{ 9\epsilon_m \omega_D }{8\pi(\epsilon_{d}+2\epsilon_m)}F, 
\end{equation}
and
\begin{equation}\label{eq_lambq}
\Lambda_Q=\left(\frac{3}{4\pi}\right)^{5/3}\frac{ 5\epsilon_m \omega_Q }{12(\epsilon_{d}+(3/2)\epsilon_m)}F^{5/3},  
\end{equation}
where $F=4\pi\rho^3/3V_{uc}$ is the metal fill fraction, {\it i.e.}, the fraction of the unit volume cell that is filled by metal.
The expressions for $\Lambda_D$ and $\Lambda_Q$ allow a first estimate of the importance of the dipole and quadrupole contributions to the plasmon-polariton dispersion and light-matter coupling. 
The ratio between the quadrupole and dipole coupling factors
\begin{equation}
\frac{\Lambda_Q}{\Lambda_D}=\frac{5}{18}\left(\frac{3}{4\pi}\right)^{2/3}\left(\frac{\epsilon_d+2\epsilon_m}{\epsilon_d+(3/2)\epsilon_m}\right)^{3/2}F^{2/3},
\end{equation}
scales with $F^{2/3}$. The prefactor ranges from $\sim 0.10$ for $\epsilon_d \gg \epsilon_m$ to $\sim 0.16$ for $\epsilon_m \gg \epsilon_d$.  In simple crystals (Bravais lattices) of spherical nanoparticles $F \leq 0.74$, with the largest packing density for FCC and HCP lattices. The quadrupole-quadrupole (QQ) interaction is in this case limited to about 13$\%$ of the dipole-dipole (DD) interaction and the dipole-quadrupole (DQ) interaction to 36$\%$. Especially, for smaller packing fractions the dipole-derived terms are expected to dominate the polariton dispersion, but we expect important contributions of the quadrupole terms for high packing. Even larger metal fill fractions may be obtained with non-spherical nanoparticles and in supercrystals with more than one nanoparticle per unit cell.\cite{Murray1980,Coropceanu2019}

The Rabi frequency associated to the interaction of light with the dipole and quadrupole plasmons can be estimated as $\Omega_R^\nu=\omega_{\bar\nu}\xi_0^\nu(\omega_{\bar\nu}/c)$ where $\bar\nu=D$, $Q$ for dipoles and quadrupoles, respectively. Within the quasi-static approximation we obtain the explicit expressions
\begin{equation}\label{Eq_OmD}
\Omega_{R}^D=\omega_D\sqrt{\frac{3F_0}{4}\left(\frac{3\epsilon_m}{\epsilon_d+2\epsilon_m}\right)}f^{1/2},    
\end{equation}
and
\begin{equation}\label{Eq_OmQ}
\Omega_{R}^Q=\omega_Q\sqrt{\frac{\pi}{3}\left(\frac{3F_0}{4\pi}\right)^{5/3}\left(\frac{5\epsilon_m}{2\epsilon_d+3\epsilon_m}\right)}f^{5/6},    
\end{equation}
where $f=F/F_0$ and $F_0$ is the maximum fill factor for a given lattice. For an FCC supercrystal the Rabi frequencies are limited to $\Omega_{R}^D=0.91\omega_D$ and $\Omega_{R}^Q=0.31\omega_Q$, which is obtained by setting $\epsilon_m \gg \epsilon_d$, $f=1$, and $F_0=0.74$ in Eqs.\ (\ref{Eq_OmD}) and (\ref{Eq_OmQ}). This places the Rabi frequencies on the order of eV for high packing densities, in excellent agreement with our recent experimental results.\cite{Mueller2020}

\subsection{Validating the model}

Before discussing and analyzing the bandstructure and properties of plasmon polaritons, we demonstrate the validity of our model by comparing it to FDTD simulations. We first describe the parameters used in both simulations. We considered an FCC nanoparticle crystal and the high-symmetry $\Gamma L$ and $\Gamma K$ directions. The nanoparticle diameters were $d=50\,$nm with interparticle (center to center) distance of $a=65\,$nm, which yields a metal fill fraction $f=0.46$. The calculations were done with the Drude model using a plasma frequency $\hbar \omega_p=9\,$eV and $\epsilon_d=1$. The nanoparticles were placed in vacuum ($\epsilon_m=1$). 

For the microscopic quantum calculation we considered Umklapp processes with $\vec G$ within up to six Brillouin zones. Plasmon-polariton energy differences of up to 10\% were obtained for some of the modes if Umklapp processes were neglected, see Fig.\,S1 for details.
The lattice vector summation in real space for calculating $S^{DD}$ was performed for $|\vec R|$ below a cutoff radius $R_{D}=60\bar R$. For $S^{DQ}$ and $S^{QQ}$ a cutoff radius of $R_Q=7\bar R$ sufficed to achieve full convergence. This reflects the fact that the dipole-quadrupole and quadrupole-quadrupole interactions fall off faster with  distance than dipole-dipole coupling. A cutoff wavevector $q_c=0.3\pi/a$ was used for the lattice sums. These parameters were used throughout the paper, unless otherwise stated. 

The FDTD simulations were done with the commercial software package Lumerical FDTD Solutions. We constructed the unit cell of an FCC crystal that is composed of spherical nanoparticles. The nanoparticles were assigned the dielectric function $\epsilon(\omega)=\epsilon_d-\omega_p^2/(\omega^2-i\gamma \omega)$ with a loss rate $\hbar \gamma = 65$\,meV (see above for the other parameters). We used a mesh size of 1\,nm to discretize space. To calculate the polariton dispersion we placed local emitters and point monitors inside the crystal.\cite{SunLin2018} We used point dipoles as light sources that radiated along the $[111]$ ($\Gamma L$) or $[110]$ ($\Gamma K$) direction. A 0.7\,fs light-pulse was injected and the electric field  recorded in the time interval from 10 to 50\,fs by a point monitor. The frequency dependent electric field was obtained by a Fourier transformation. We used Bloch periodic boundary conditions to choose a specific wave vector. By running a sweep of simulations for different wave vectors we obtained the polaritonic band structure.

       \begin{figure}[!htb]
        \centering
        \includegraphics[width=8.5cm]{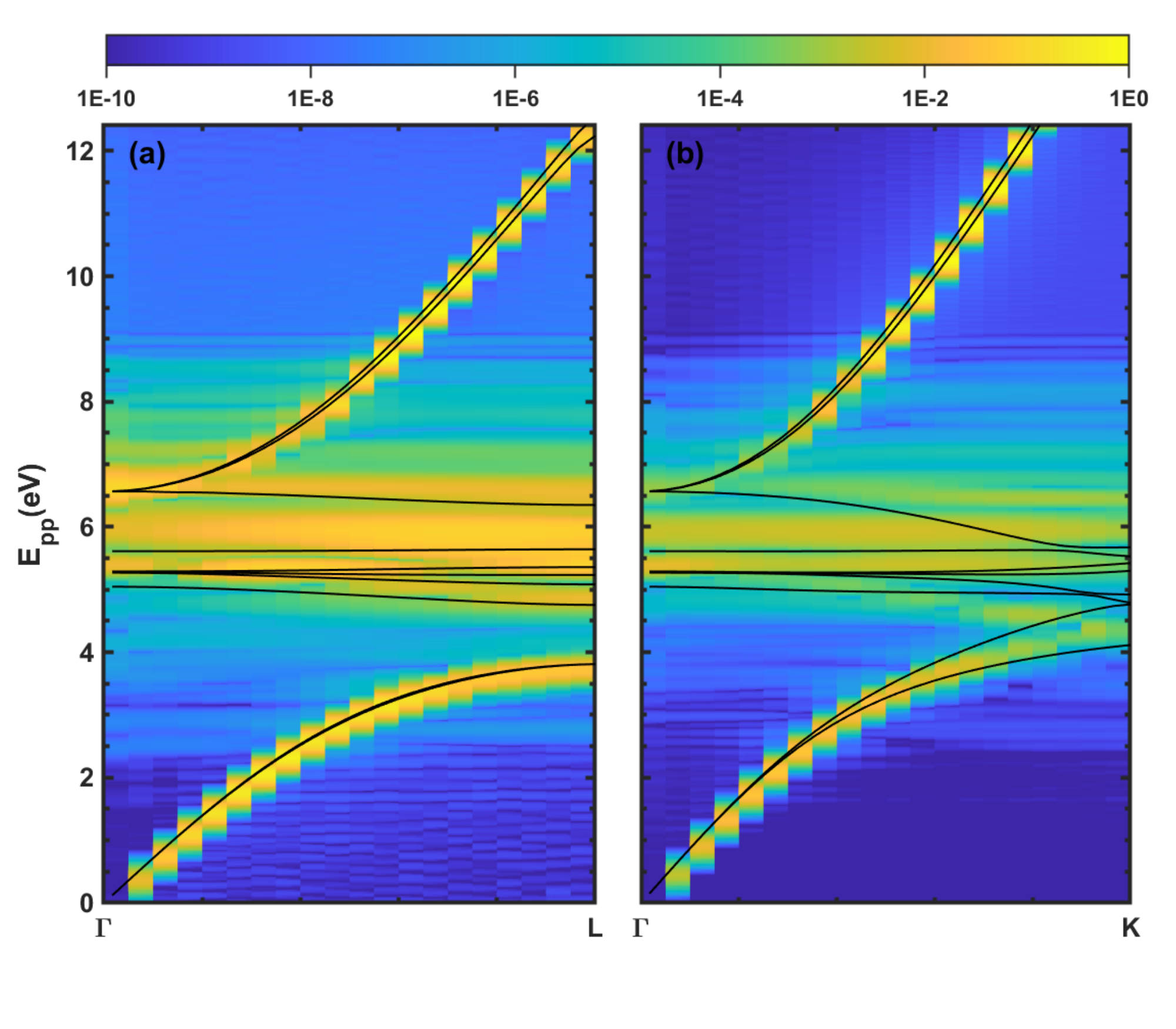}
        \caption{Polariton band structure of an FCC crystal of spherical nanoparticles along the (a) $\Gamma L$ and (b) $\Gamma K$ high-symmetry directions. Full lines were calculated with the microscopic quantum model.  The colormap shows the magnitude (in log scale) of the electric field obtained in FDTD simulations as a function of energy and momentum. $\hbar \omega_p=9$ eV, $\epsilon_m=\epsilon_d=1$, $a=65$ nm and $f=0.46$.}
        \label{fig:FDTD_MQPP}
        \end{figure}

Figure \ref{fig:FDTD_MQPP} compares the band structure obtained with FDTD and the microscopic model. The background of the figure is a color map of the integrated intensity of the electric field as a function of $\omega$ and $q$, which corresponds to the polariton dispersion predicted by the FDTD simulations. The black lines show the plasmon-polariton dispersion calculated with the microscopic quantum model. Our model reproduces the FDTD dispersion very well. The far-field response of the supercrystal is dominated by the two dipole-derived bands.\cite{Mueller2020} These are the bands with lowest and highest energy in Fig.~\ref{fig:FDTD_MQPP}, which are excellently described by the quantum mechanical model. Along the $\Gamma L$ direction, Fig.~\ref{fig:FDTD_MQPP}(a), the quadrupole bands between the two dipole-derived states agree also between FDTD and the microscopic model. The FDTD simulation appears to contain more states, which originate from hexapole eigenmodes of the nanoparticles or artefacts of the simulation. 
The $\Gamma K$ direction, Fig.~\ref{fig:FDTD_MQPP}(b), is the high-symmetry direction  of the FCC lattice that  is most strongly affected by the quadrupole modes. As discussed below, the dipole-only model strongly overestimates the energy of the lowest lying polariton band near the $K$ point, whereas the inclusion of the quadrupole modes results in pretty good agreement with the FDTD results. 

The two FDTD simulations shown in Fig.~\ref{fig:FDTD_MQPP} took several hours each, whereas the microscopic quantum mechanical band structure was obtained in seconds. Our model allows a rapid screening of many supercrystal structures, fill factors, nanoparticle shapes, and so forth. Its true strength, however, goes beyond its computational capability: The microscopic model allows an in-depth study of the origin of the plasmon-polariton band structure and its properties as we will show in the following section.

\section{Results and discussion}
\label{sec:results}

We modeled the plasmon-polariton band structure of FCC nanoparticle supercrystals using our microscopic model. 
With the simulations we can explain the contribution of the interaction between the nanoparticles and with the electromagnetic modes to the final polariton states. We are able to extract the coupling and mixing of dipole- and quadrupole-derived states in this particular Bravais lattice. Finally, we show how to extract the dipolar, quadrupolar, and photonic contribution to each polariton state. The results impressively reproduce the decoupling of light and matter in the USC and DSC regime.\cite{DeLiberato2014,Mueller2020}

\subsection{Collective plasmon modes}

        \begin{figure*}[!htb]
        \centering
        \includegraphics[width=17.5cm]{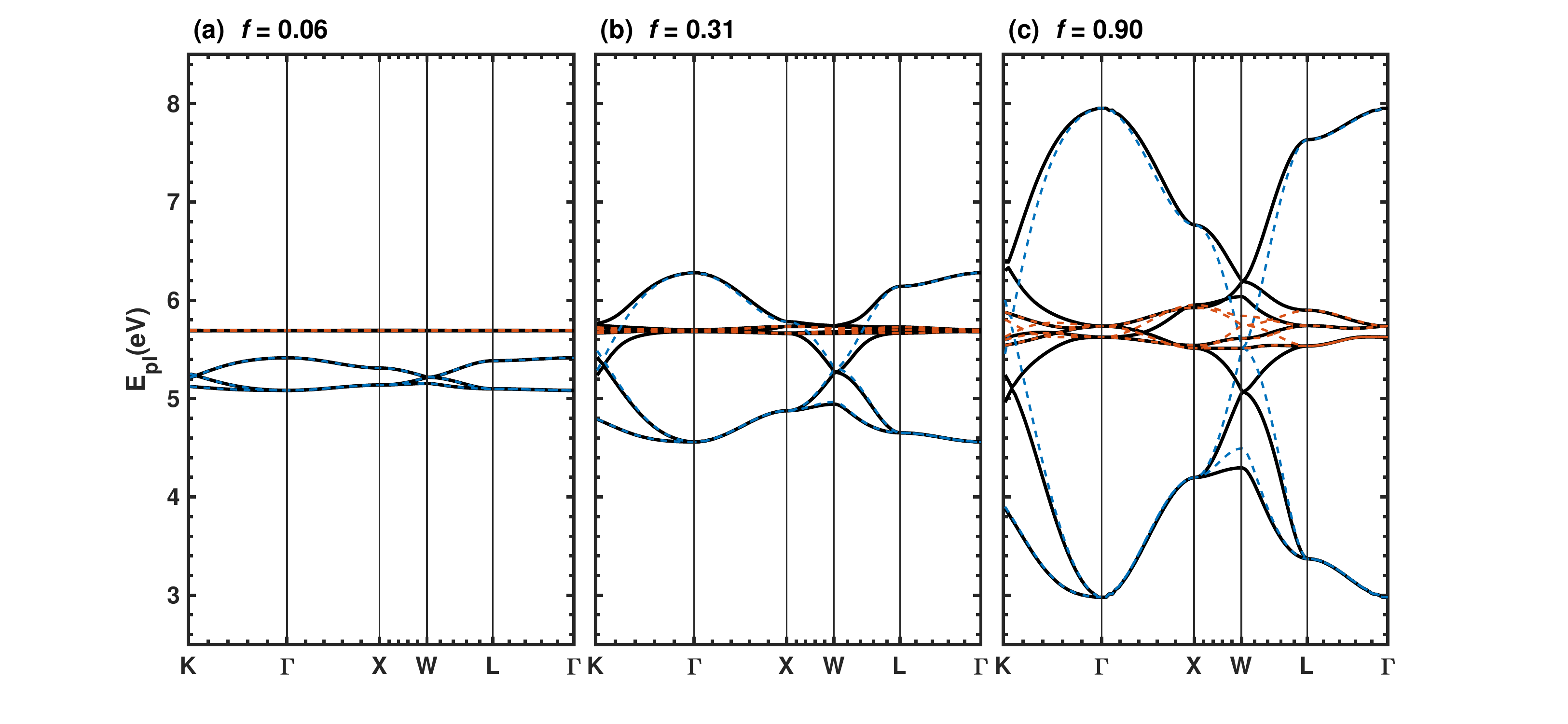}
        \caption{Collective plasmon dispersion along the high symmetry directions of an FCC lattice for fill factors  (a) $f=0.06$, (b) $0.31$, and (c) $0.90$. The blue (red) lines are induced by the dipole (quadrupole) plasmons of the nanoparticles neglecting dipole-quadrupole interactions. The black lines are a full calculation including dipole and quadrupole modes and  their interaction. $\hbar \omega_p=9$ eV, $\epsilon_m=\epsilon_d=1$ and $a=62$~nm.}
        \label{fig:PlasmonicBands}
    \end{figure*}

We model the optical properties of plasmonic nanoparticle supercrystals in a step-by-step approach. We start with the interaction between dipole and quadrupole nanoparticle states that give rise to collective plasmon modes. This initial  plasmonic band structure omits the coupling to electromagnetic states.\cite{Weick2015,Lamowski2018, Mueller2020} Including the photons will later create the supercrystal plasmon polaritons. 

In Fig.~\ref{fig:PlasmonicBands} we show the plasmonic bandstructure of an FCC crystal considering both dipole and quadrupole nanoparticle excitations. In each panel we also show the bandstructures for the dipole (blue) and quadrupole (red) modes when turning off the interaction between the  dipole and quadrupole states. For the lowest fill factor $f=0.06$ in Fig.~\ref{fig:PlasmonicBands}(a) the dipole and quadrupole states do not cross and are largely decoupled as can be seen by the agreement between the black and the blue/red lines.  The lowest plasmonic state at the $\Gamma$ point [$\sim 5.1$ eV in Fig. \ref{fig:PlasmonicBands}(a)] is a two-fold degenerate dipole-induced state.\cite{Lamowski2018} It remains degenerate along the $\Gamma X$ and $\Gamma L$ directions but splits along $\Gamma K$. These two bands are associated with transverse oscillations of the plasmons, {\it i.e.}, the electrons oscillate perpendicular to the propagation direction. The uppermost dipole-induced band [$\sim 5.4$ eV at $\Gamma$ in Fig.~\ref{fig:PlasmonicBands}(a)] is a longitudinal oscillation that does not couple directly with light. 

The quadrupole states are constant across the Brillouin zone for $f=0.06$, but become dispersive  for the larger fill factors $f = 0.31$, Fig.~\ref{fig:PlasmonicBands}(b), and $0.90$, Fig.~\ref{fig:PlasmonicBands}(c). The quadrupole states consist of five bands that are two- and three-fold degenerate at the $\Gamma$ point [for example, at $5.62\,$ and $5.75\,$eV in Fig.~\ref{fig:PlasmonicBands}(c)]. Along the $\Gamma L$ direction the three-fold degenerate state splits into a two-fold and a non-degenerate band, while the lower branch remains two-fold degenerate. Along the high-symmetry lines the bands split and cross, but overall the quadrupole dispersion is much narrower (0.3 eV for $f=0.9$) than the dispersion of the dipole bands (5 eV for for $f=0.9$). The reason is that the dipole-dipole coupling is much stronger than the quadrupole-quadrupole coupling, with a ratio of $\Lambda_D/\Lambda_Q\sim 13$.

The dispersion of the dipole-derived plasmon band increases rapidly with metal fill fraction, Fig.~\ref{fig:PlasmonicBands}. 
For $f>0.1$ the dipole band cross the energy of the  quadrupole states. The two types of bands overlap and the dipole-quadrupole interaction strongly affects the plasmonic dispersion. The magnitude of this interaction can be qualitatively evaluated by observing the differences between the black lines (including $DQ$ interaction) and the blue and red lines in Fig.~\ref{fig:PlasmonicBands}. For $f=0.31$ and $0.90$ the differences is very pronounced, especially in the $\Gamma K$ and the $XWL$ directions. This is in contrast with the results for $f=0.06$, Fig.~\ref{fig:PlasmonicBands}(a), where the black and blue/red lines are identical throughout the Brillouin zone. 

The dipole-quadrupole mixing depends on the symmetry of the plasmonic bands. For example, along the $\Gamma L$ direction the states remain unchanged by the coupling, indicating that dipole and quadrupole modes cannot couple in this high-symmetry direction. This explains why the dipole approximation worked very well for analysing the optical spectra of gold nanoparticle supercrystals where the light propagated normal to the (111) surface,\cite{Mueller2020} see discussion further below. In contrast, $DQ$ coupling is allowed along the $\Gamma K$ and the $XWL$ high-symmetry lines. The mixing  prevents the crossing of the transverse and longitudinal dipole-derived bands at $W$ and close to $K$, see Fig.~\ref{fig:PlasmonicBands}. 
Dipole-quadrupole coupling also reduces the splitting of the transverse states. For the $\Gamma K$ direction, two out of the five quadrupole bands are strongly mixed with the dipole modes, while the other three remain practically unchanged. We also note that the lowest transverse dipole bands are less affected by the dipole-quadrupole coupling, because of the larger energy difference between the states.

\subsection{Plasmon polaritons}      

        \begin{figure*}[!htb]
        \centering
        \includegraphics[width=17.5cm]{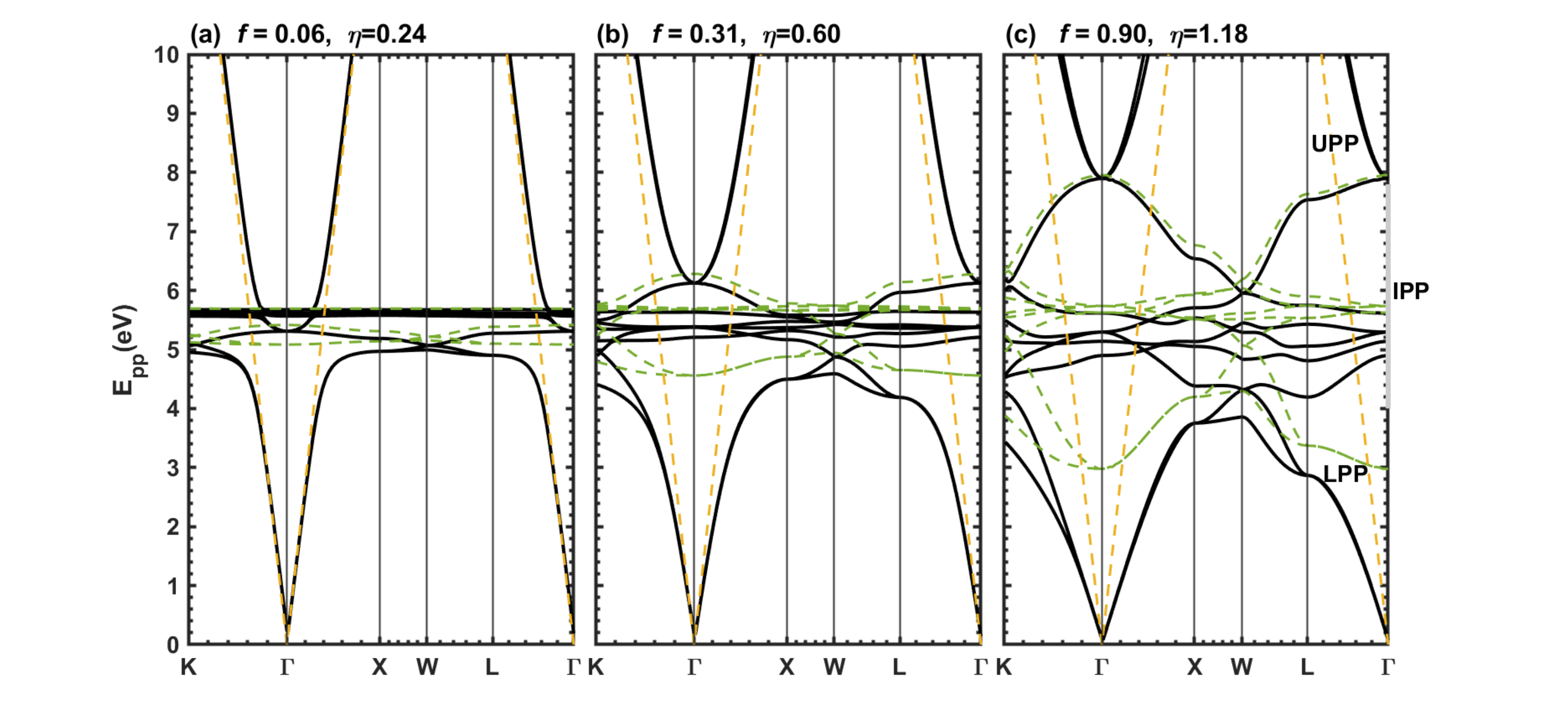}
        \caption{Plasmon-polariton dispersion along the high symmetry directions of an FCC crystal with metal fill factors (a) $f=0.06$, (b) $0.31$, and (c) $0.90$. Black lines are calculated including all terms and interactions, while green dashed lines show the dispersion of the bare plasmons. Yellow dashed lines show the bare photon dispersion. $\hbar \omega_p=9$ eV, $\epsilon_m=\epsilon_d=1$ and $a=62$ nm.}
        \label{fig:Plasmon_Polariton}
    \end{figure*}

After having examined the bare plasmon bands, we include the coupling to free-space photons and calculate the plasmon-polariton dispersion. The polaritons are coupled electronic and electromagnetic eigenstates of the nanoparticle supercrystals.\cite{LiJensen2006, HuangChengPing2010, Lamowski2018, Mueller2020} In experiments with gold nanoparticles supercrystals these excitations determined the optical response for energies below the interband transitions.\cite{Mueller2020} Figures \ref{fig:Plasmon_Polariton}(a)-(c) show the plasmon-polariton bands obtained by including the plasmon-photon interaction, 
considering Umklapp processes up to the sixth Brillouin zone ($n_{BZ}=6$). To allow for a comparison, we also show as dashed lines the bare plasmon (black) and photon (yellow) energies. For the smallest fill fraction ($f=0.06$), light-matter interaction is determined mainly by the dipole excitations. The coupling of the transverse dipole bands and photons gives rise to a pronounced level anticrossing, resulting in plasmon-polariton bands with a dispersion $E_{pp}$ that is very different from the uncoupled states.\cite{LiJensen2006, HuangChengPing2010, Lamowski2018, Mueller2020} We observe two nearly degenerate parabolic bands centered at the $\Gamma$ point, which we will refer to as the upper plasmon-polariton (UPP) and two lower bands, the lower plasmon-polaritons (LPP). The LPPs start off as linear bands with vanishing energy at the $\Gamma$ point. They bend down and become almost flat at the zone edges. The quadrupole bands remain flat over the entire Brillouin zone in Fig.~\ref{fig:Plasmon_Polariton}(a), because the interaction with light is negligible at this metal fill fraction. The longitudinal dipole-derived bands do not couple with light and their polariton dispersion remains unchanged compared to the bare plasmons.

With increasing metal fill factor quadrupole modes mix with the dipole plasmons and the photons resulting in six intermediate plasmon-polaritons (IPPs), Fig.~\ref{fig:Plasmon_Polariton}(b) and (c). This occurs because the coupling between the plasmon and between plasmons and photons increases with metal fill fraction. The topmost IPP band corresponds mainly to the longitudinal dipole-derived band that does not couple directly with light and only weakly with the quadrupole modes. The five other bands are mainly composed of quadrupole-like oscillations which are downshifted by their interaction with the electromagnetic field. This downshift is different for each of the bands, effectively increasing the bandwidth of IPPs. For instance, for $f=0.90$, the energy difference between the lower and upper quadrupole-derived plasmon-polaritons at the $X$ point is $\sim 1.3$ eV. This is more than four times larger than for the plasmon bands and no coupling to photons, width $0.3$\,eV see dashed lines.    

The UPP and LPP bands in the $\Gamma L$  direction are derived from the dipole modes without quadrupole mixing, due to the absence of dipole-quadrupole interaction along this high-symmetry direction. This explains why the polariton band structure observed along the $[111]$ direction of gold nanoparticle supercrystals was excellently described by a single band Hopfield model and a microscopic calculations within the dipole approximation\cite{Mueller2020}. The situation is different along the $\Gamma K$ and the $XWL$ directions, for which the  dipole-quadrupole coupling is strong, as seen in Fig.~\ref{fig:PlasmonicBands}. Along these directions, the dipole-quadrupole and the light-quadrupole interactions lead to an anti-crossing between the quadrupole and dipole bands, thus effectively pushing down the topmost LPP. This result shows that a complete and accurate description of metallic supercrystals requires that quadrupole modes and Umklapp processes, see Supplementary Information, are included in the model. The strong dependence of the coupling on the direction in the Brillouin zone points towards symmetry-based selection rules for dipole-quadrupole and light-matter coupling, which would be interesting to study for various crystal symmetries.

\begin{figure}[!htb]
        \centering
        \includegraphics[width=12cm]{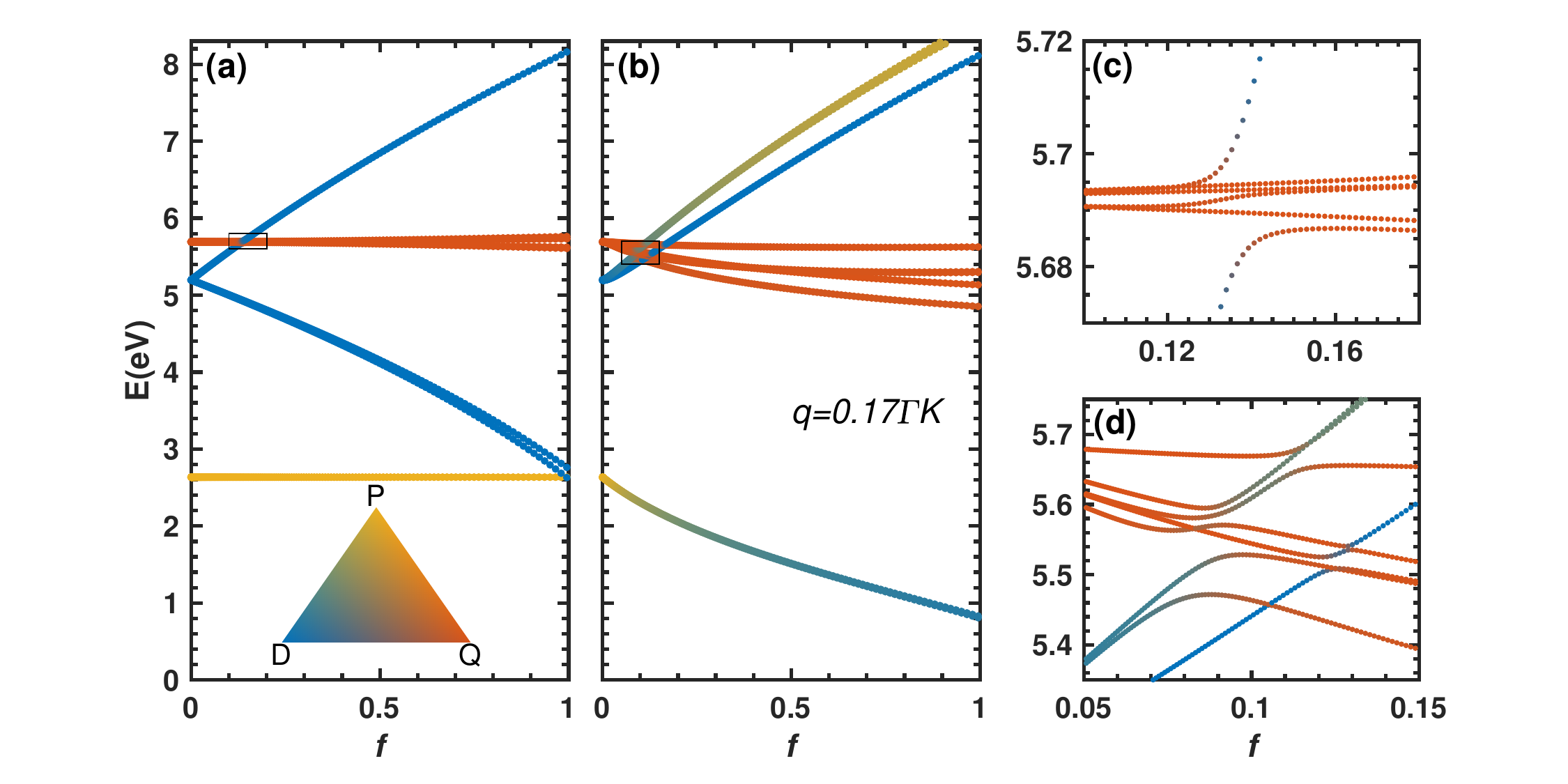}
        \caption{Energies of (a) the bare plasmon and photon states and of (b) the plasmon-polariton states for $q=0.17\Gamma K$. The coloring indicates the dipole, quadrupole, and photon contribution to each state - see inset in (a). (c) and (d) show zoomed images of the  rectangular areas in (a) and (b), respectively.}
        \label{fig:WxF}
    \end{figure}

We now examine the coupling and level anticrossing of the dipole and quadrupole modes and the photons in greater detail. In Fig. \ref{fig:WxF}(a) we show the energies of the bare plasmon and photon bands at $q=0.17\Gamma K$ as a function of the metal fill factor and in Fig. \ref{fig:WxF}(b) the corresponding energies of the plasmon-polariton bands. The colors indicate the magnitude of the contribution of dipole (blue) and quadrupole (red) modes and  photons (yellow) to the states according to the color code triangle in (a). Without light-matter coupling, the longitudinal dipole-derived plasmon modes [blue lines in Fig.~\ref{fig:WxF}(a)] upshift almost linearly with filling, while the quadrupole energies remain nearly constant. At $f\sim 0.12$, the energies of the two sets of bands cross. For one of the quadrupole modes, the interaction with the longitudinal dipole band causes an avoided crossing  with a gap on the order of $0.01\,$eV, see the enlarged panel in Fig.~\ref{fig:WxF}(c). The other quadrupole and dipole bands are only weakly affected by the $DQ$ interaction. 

Without light-matter coupling the photon energy remains at $E=2.6$\,eV independent of filling. When light-matter coupling is "turned on", plasmon polaritons form. The UPP band is mainly composed of transverse dipole-derived plasmon states at $0.17\,\Gamma K$ and for vanishing metal content $f\sim 0$ . With increasing filling the UPPs become more photon-like (yellow color); at the same time, their energy increases in parallel with the longitudinal dipole mode (blue) that does not interact with light. The LPP shifts to smaller energies with increasing $f$ and obtains a strong dipole plasmon contribution. 

The spectral range of the anticrossing of the UPPs and the quadrupole bands is shown at higher magnification in Fig.~\ref{fig:WxF}(d). Along the $\Gamma K$ direction, dipole plasmons, quadrupole plasmons, and photon all mix into polariton states. As the polarization dependence of the light-quadrupole and light-dipole interactions are different, we expect cross-polarized absorption and chiral activity, which will be studied in a future work. As the fill factor increases further, the UPP bands become increasingly photon-like and their interaction with the quadrupole modes causes the latter to downshift in energy, thus increasing the overall bandwidth of the quadrupole modes. 

\subsection{Normalized coupling strength}

The normalized coupling strength $\eta=\Omega_R/\omega_0$ compares the Rabi frequency $\Omega_R$ to the bare frequency $\omega_0$ of the system. $\Omega_R$ can be found from the minimum energy splitting between the LPP and UPP bands (divided by two) and $\omega_0$ from a calculation of the bare plasmon dispersion.\cite{Baranov2020,Kockum2019,FornDiaz2019} In this section we derive a close expression for $\eta$ as a function of our model parameters. It will facilitate chosing a nanoparticle supercrystal for a desired coupling. We will show that a wide range of USC and DSC light-matter interaction can be realized in plasmonic supercrystals.

\begin{figure}[!htb]
        \centering
        \includegraphics[width=8cm]{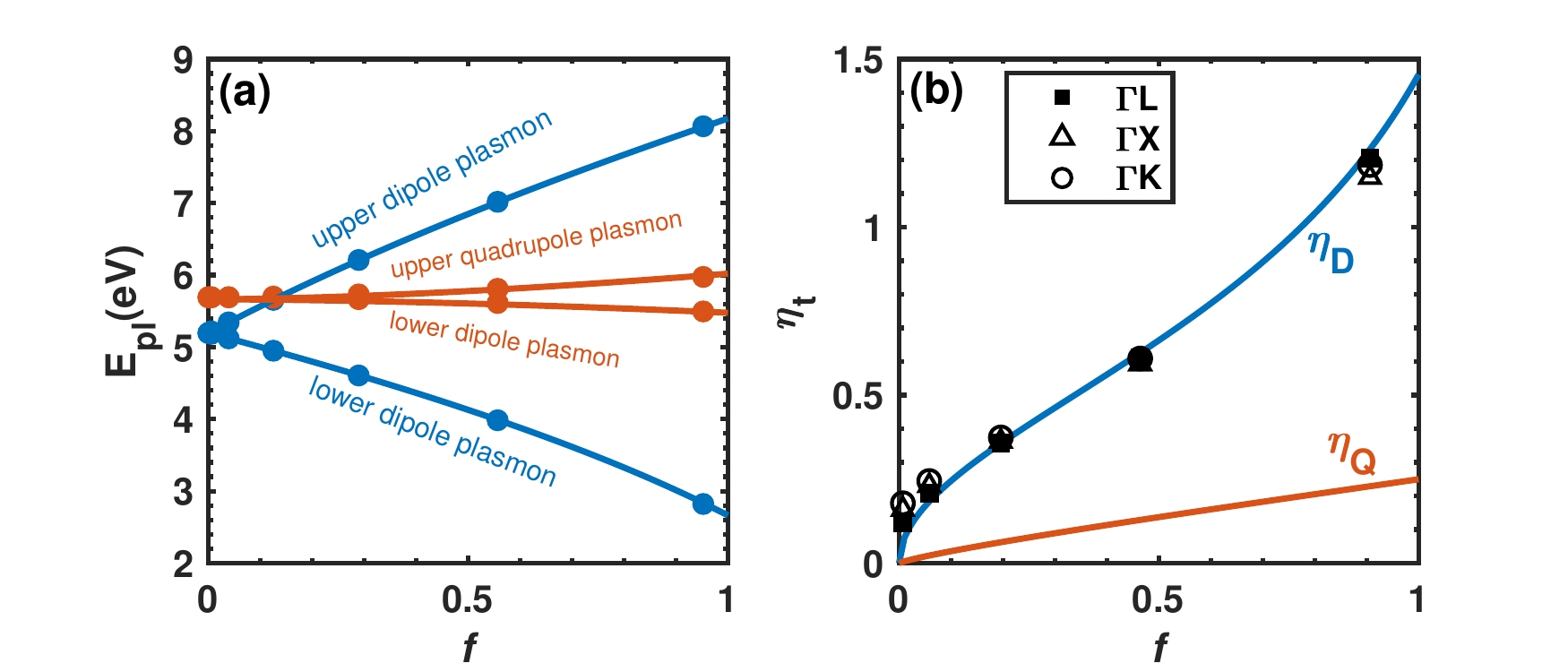}
        \caption{(a) Energies of the upper and lower dipole (blue symbols) and quadrupole (red symbols) plasmon bands for different values of $f$  of an FCC crystal. The dipole (quadrupole) energies are evaluated at the $\Gamma$ ($X$) point. Red and blue lines are fits to the data points, using Eq.~\ref{eq_omegapl}, see Table\ \ref{tab:etalamb}. (b) Normalized coupling strength $\eta_t$ of the transverse dipole-derived plasmon band for the $\Gamma L$ (squares), $\Gamma X$ (triangles) and $\Gamma K$ (dots) high-symmetry directions. Blue and red lines are coupling strengths predicted with Eq. (\ref{eq_eta}) for the dipole and quadrupole modes, respectively.}
        \label{fig:etaCalc}
    \end{figure}

The bare plasmon energies without coupling to the electromagnetic states depends on the plasmon coupling factors $\Lambda_{\bar \nu}$ ($\bar\nu=D,Q$) defined in Eqs. (\ref{eq_lambd}) and (\ref{eq_lambq}). The energies are well described by\cite{Mueller2020}
\begin{equation}\label{eq_omegapl}
    E_{pl,\nu}(\vec q)=\hbar\omega_{\bar\nu}\sqrt{1+ s_{\nu,f}(q)\Lambda_{\bar \nu} },
\end{equation}
where $s_{\nu,f}(q)$ incorporates the effects of the lattice; it depends only weakly on $f$. $s_{\nu,f}(q)$ measures the enhancement of the effective plasmon-plasmon coupling due to the crystalline structure. This value can be calculated numerically for each of the plasmon bands at any given $q$.\cite{Coropceanu2019} However, as we are interested in an effective expression, we use a different approach and fit the energy of the lowest lying $D$ and $Q$ bands at the $\Gamma$ and $X$ points, respectively. These are the wavevectors of the largest bandwidths of the dipole- and  quadrupole-derived bands, which gives us an overall measure for plasmon-plasmon coupling.  Figure~\ref{fig:etaCalc}(a) shows the energies of the upper and lower dipole and quadrupole bands at the $\Gamma$ and $X$ point. These are fitted to Eq.(\ref{eq_omegapl}) and the fitting parameters are shown in Table \ref{tab:etalamb}. 

\begin{table}[]
    \centering
    \begin{tabular}{cccc}\hline\hline
          &  label & $q$ & value \\\hline 
        upper dipole & $s_D^u$ & $\Gamma$ & 16.5  \\
        lower dipole & $s_D^l$ & $\Gamma$ & -8.3  \\
        upper quadrupole & $s_Q^u$ & $X$ & 14  \\
        lower quadrupole & $s_Q^l$ & $X$ & -7\\\hline\hline
    \end{tabular}
    \caption{Fit parameters for the upper and lower dipole and quadrupolar states at $q=\Gamma$ and $X$, respectively. Calculated results were fitted to Eq.(\ref{eq_eta}) with one free parameter.}
    \label{tab:etalamb}
\end{table}

We now use the Rabi frequencies and dipole and quadrupole coupling strengths in Eqs.\ \eqref{Eq_OmD}-\eqref{Eq_OmQ} and combine it with the plasmon energy in Eq.~(\ref{eq_eta}). We find a compact expression for 
the maximum reduced coupling 
\begin{equation}\label{eq_eta}
    \eta_\nu=\sqrt{\frac{2\pi\Lambda_{\bar \nu}}{(1+\bar s^l_\nu\Lambda_{\bar \nu})}},
\end{equation}
where $s^l_\nu$ are the fitting parameters for the lowest energy dipole and quadrupole bands at the chosen points, see Table \ref{tab:etalamb}. For a single plasmonic state $\eta$ can be found from Eq.~\eqref{eq_eta}. A more general expression is necessary for mixed dipole and quadrupole bands. As we are mainly interested in the maximum normalized coupling strength, we will focus on the coupling to the lowest energy dipole- and quadrupole-derived bands.
The solid lines in Fig.~\ref{fig:etaCalc}(b) show the dependence of the reduced dipole and quadrupole coupling in Eq.~\eqref{eq_eta} on the fill fraction. The symbols are the reduced coupling strengths of the transverse dipole-derived bands $\eta_t=\Delta E_\mathrm{UL}/(2E^t_{pl,D})$ from a microscopic quantum calculation evaluated at the crossing $|q_0|\sim E^t_{pl,D}/\hbar c$ of the bare dipole plasmon and the photon dispersion along $\Gamma L$, $\Gamma X$ and $\Gamma K$. $E^t_{pl,D}$ is the energy of the transverse dipole-induced plasmon and $\Delta E_{UL}\sim E_\mathrm{UPP}(q_0)-E_\mathrm{LPP}(q_0)$ is the energy difference between the upper and lower plasmon polariton branches. The normalized coupling strength obtained for the different  directions correspond well with the value obtained by the expression for the dipole coupling strentgh $\eta\sim \eta_D$. The fact that this is true even for the $\Gamma K$ direction indicates that the quadrupole contribution the coupling strength of the transverse dipole is negligible. Furthermore, it shows that Eq.\ \eqref{eq_eta}, with the parameters in Table\ \ref{tab:etalamb}, can be used to estimate the maximum coupling strength in FCC supercrystals. With this, for a metal fill fraction of 3\% the maximum coupling strength is on the order of $\eta=0.13$, and therefore already in the USC regime ($\eta>0.1$), while the DSC regime is reached for fill fractions above 80\%, Fig.~\ref{fig:etaCalc}(b). 

\subsection{Decoupling of light and matter the USC and DSC regimes}\label{sec:insights}

        \begin{figure*}[!htb]
        \centering
        \includegraphics[width=17cm]{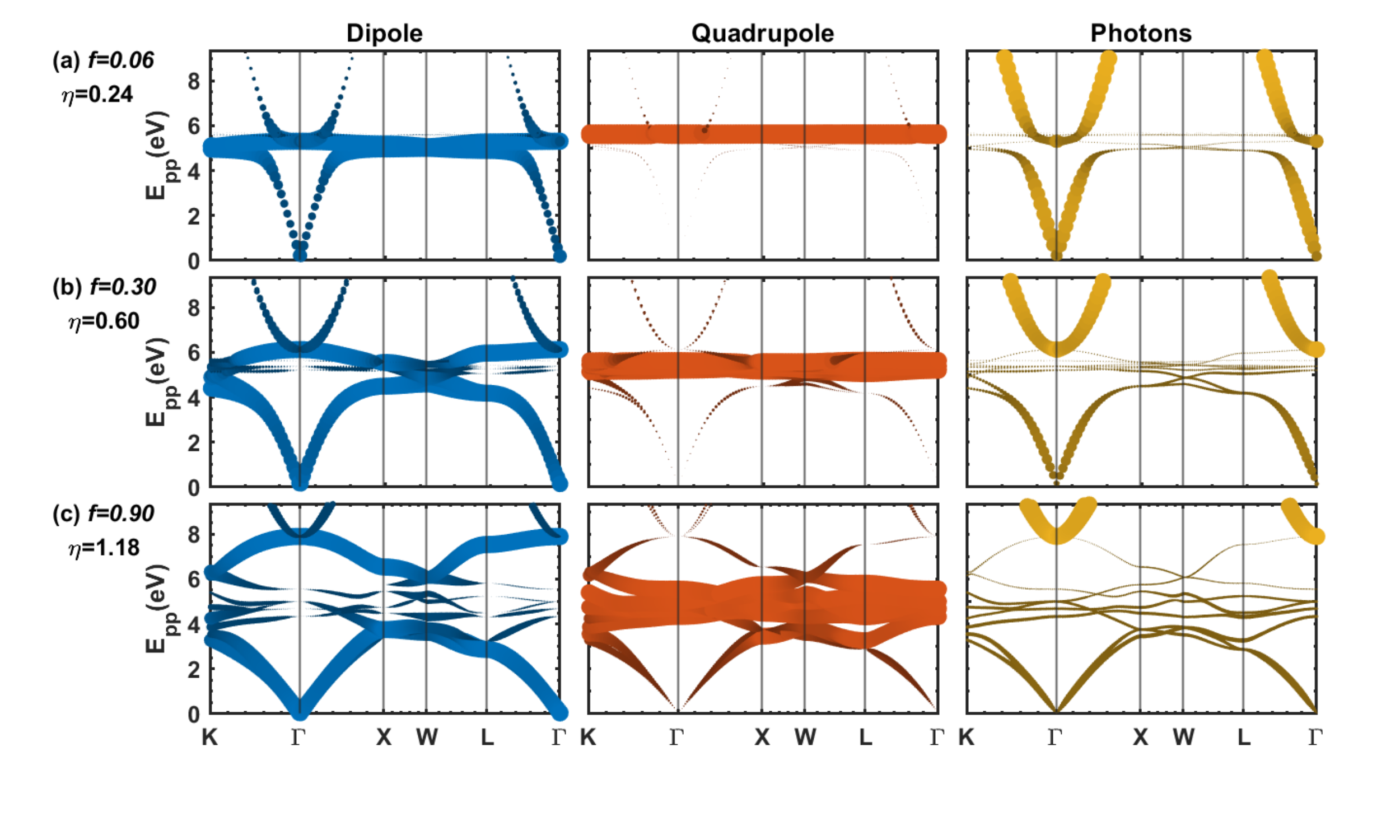}
        \caption{Decomposition of the plasmon-polariton bands into their dipole (blue), quadupole (red) and photon (yellow) contributions for fill factors (a)$f=$ 0.06,(b) 0.31, and (c) 0.90. The size of the data points shows the magnitude of the relative contributions of each of these bare excitations to the plasmon-polariton bands.}
        \label{fig:decomp}
    \end{figure*}

A fascinating signature of ultrastrong and deep strong coupling is the decoupling of light and matter in space and in frequency that leads to a breakdown of the Purcell effect.\cite{DeLiberato2014} The Purcell effect describes the increase in radiative damping with increasing light-matter coupling.\cite{Purcell1946} In the weak and strong coupling regimes ($\eta\ll 1$) the radiative damping scales with $\eta^2$. De Liberato \cite{DeLiberato2014} predicted that the Purcell effect saturates around $\eta\sim 0.5$ and radiative damping decreases for higher values of $\eta$. Mueller {\it et al.} \cite{Mueller2020} demonstrated the breakdown of the Purcell effect in plasmonic supercrystals for $\eta > 1$.  Although this breakdown can be described classically,\cite{DeLiberato2014}  the microscopic quantum description of the plasmon-polaritons allows us to directly observe the decoupling  with increasing fill fraction, {\it i.e.}, increasing light-matter coupling. 

We use the transformation matrix $T_{\vec q}$ introduced in Sect.~\ref{sec:NPcrystals} to decompose the plasmon-polariton states into the bare dipole and quadrupole plasmonic and the photonic components. In Fig.~\ref{fig:decomp} we show the  plasmon-polariton states decomposed into dipole plasmons, quadrupole plasmons, and photons for three fill factors. For small filling $f=0.06$ the plasmonic characters are mainly localized in the energy regions of the dipole and quadrupole plasmon bands. The band with linear dispersion are  photonic in character. There is no mixing between the dipole and quadrupole states as expected from our discussion of the bare plasmon dispersion. The dipole plasmon and the photon mix slightly at their crossing, so that the linear bands show a small dipole-plasmon character while the flat band at about $5.19\,$eV shows a weak photonic character. 

With increasing $f$ the mixing between the three states becomes more apparent. For $f=0.31$ the dipole plasmonic character of the linearly dispersive bands is very pronounced. We also see a non-negligible mixing between the dipole and quadrupole states as well as the quadrupole modes and photons. Peaks related to quadrupole modes should start to appear in the absorption spectra and affect the overall dispersion of the polaritons. The top plasmon-polariton bands acquires a finite mass, because it is composed of photons, dipole plasmons and even a small contribution from quadrupole modes. Finally, the distribution of the plasmon and photon states becomes asymmetric for lower and upper polaritons: While the low-energy states have a stronger plasmonic component, the upper polaritons are photonic in character.

For the high metal fill fraction $f = 0.9$, the linear bands starting at zero energy are almost entirely composed of dipole states, having only weak photonic and quadrupole plasmonic character. The weakly dispersive bands in the gap between UPP and LPP remain strongly plasmonic in nature and are predominantly composed of quadrupole modes, but the states became  mixed with photons and with dipole plasmons. This indicates that they should be accessible optically. The UPP branches developed into a pair of massive bands mainly composed of photons. Overall, it is striking that there is little overlap between the three different types of quasi-particles, which is a manifestation of the light-matter decoupling in the DSC regime.\cite{DeLiberato2014,Mueller2020} Indeed, the distribution of states for highest filling resembles the low-filling case: Each component - dipole, quadrupole, and photon - is concentrated in a portion of the polariton dispersion with little contribution to the other states. 

The quantum model proposed here gives insight into the nature of the plasmon-polariton states in addition to its excellent  description of the plasmon-polariton band structure. In future, it may be applied to calculate quantum related properties such as the squeezing of plasmons and photons, the population of the supercrystal ground state with photons and plasmons, and correlation functions of the bare excitations. \cite{Artoni1991,Ciuti2005, Cirio2016,Kockum2019,FornDiaz2019}

\section{Conclusions}
\label{sec:conclusions}

In conclusion, we proposed a microscopic model to calculate plasmon polaritons in nanoparticle supercrystals. Our model includes the dipole and quadrupole modes of the nanoparticle building blocks and their coupling to photons. We show that the mixing of the dipole and quadrupole-derived states is important for calculating the collective plasmon and plasmon-polariton sates. The microscopic quantum model leads to a closed expression for the reduced light-matter coupling strength of the dipole and quadrupole modes. The dipole derived states of FCC nanoparticle supercrystals are in the ultrastrong coupling regime for all realistic fill fractions and enter deep strong coupling for a fill fraction of $0.8$ (assuming vacuum between the nanoparticles in the crystal). 
The quantum based calculations give insight into the unique properties of strongly coupled systems as we show for the example light-matter decoupling in the DSC regime. The model can be applied for different lattice structures including lattices with more than one particle in the basis. It will contribute to the study and optimization of the many  supercrystal structures currently being developed. 

\section*{Acknowledgements}
\label{sec:acknowledgements}
We thank J. Weick for useful discussions. E.B.B acknowledges financial support from CNPq, CAPES (finance code 001) and FUNCAP (PRONEX PR2-0101-00006.01.00/15). S.R. and N.S.M. acknowledge support by the European Research Council ERC under grant DarkSERS (grant number 772108). This work was supported by the Focus Area NanoScale of Freie Universit\"at Berlin.

\newpage
\bibliographystyle{unsrt}
\bibliography{references_mendeley,references2}

\begin{thebibliography}{10}

\bibitem{Murray1980}
M.~J. Murray and J.~V. Sanders.
\newblock Close-packed structures of spheres of two different sizes {II. T}he
  packing densities of likely arrangements.
\newblock {\em Phil. Mag. A}, 42:721--740, 1980.

\bibitem{boles_self-assembly_2016}
Michael~A. Boles, Michael Engel, and Dmitri~V. Talapin.
\newblock Self-{Assembly} of {Colloidal} {Nanocrystals}: {From} {Intricate}
  {Structures} to {Functional} {Materials}.
\newblock {\em Chemical Reviews}, 116(18):11220--11289, 2016.

\bibitem{murray_synthesis_2000}
C.~B. Murray, C.~R. Kagan, and M.~G. Bawendi.
\newblock Synthesis and {Characterization} of {Monodisperse} {Nanocrystals} and
  {Close}-{Packed} {Nanocrystal} {Assemblies}.
\newblock {\em Annual Review of Materials Science}, 30(1):545--610, 2000.

\bibitem{Shevchenko2006}
E.~V. Shevchenko et~al.
\newblock Structural characterization of self-assembled multifunctional binary
  nanoparticle superlattices.
\newblock {\em J. Am. Chem. Soc.}, 128:3620--3637, 2006.

\bibitem{Coropceanu2019}
Igor Coropceanu, Michael~A. Boles, and Dmitri~V. Talapin.
\newblock Systematic mapping of binary nanocrystal superlattices: {T}he role of
  topology in phase selection.
\newblock {\em J. Am. Chem. Soc.}, 141:5728--5740, 2005.

\bibitem{Mueller2020}
Niclas~S. Mueller, Yu~Okamura, Bruno G.~M. Vieira, Sabrina Juergensen, Holger
  Lange, Eduardo~B. Barros, Florian Schulz, and Stephanie Reich.
\newblock {Deep strong light–matter coupling in plasmonic nanoparticle
  crystals}.
\newblock {\em Nature}, 583(7818):780--784, 2020.

\bibitem{Schulz2020}
F.~Schulz, O.~Pavelka, F.~Lehmkühler, F.~Westermeier, Y.~Okamura, N.S.
  Mueller, S.~Reich, and H.~Lange.
\newblock Structural order in plasmonic superlattices.
\newblock {\em Nat. Comm.}, 11:3821, 2020.

\bibitem{Vovk2020}
Sebastian Vovk, Nuri Yazdani, and Vanessa Wood.
\newblock Manipulating electronic structure from the bottom-up: {C}olloidal
  nanocrystal-based semiconductors.
\newblock {\em J. Phys. Chem. Lett.}, 11:9255--9264, 2011.

\bibitem{Baimuratov2013}
Anvar~S. Baimuratov, Ivan~D. Rukhlenko, Vadim~K. Turkov, Alexander~V. Baranov,
  and Anatoly~V. Fedorov.
\newblock Quantum-dot supercrystals for future nanophotonics.
\newblock {\em Sci. Rep.}, 3:1727, 2013.

\bibitem{Mueller2021}
Niclas~S. Mueller, Emanuel Pfitzner, Yu~Okamura, Georgy Gordeev, Patryk Kusch,
  Holger Lange, Joachim Heberle, Florian Schulz, and Stephanie Reich.
\newblock Surface-enhanced raman scattering and surface-enhanced infrared
  absorption by plasmon polaritons in three-dimensional nanoparticle
  supercrystals.
\newblock {\em ACS Nano}, 15(3):5523--5533, 2021.

\bibitem{GarciaLojo2019}
Daniel García-Lojo, Sara Núñez-Sánchez, Sergio Gómez-Graña, Marek
  Grzelczak, Isabel Pastoriza-Santos, Jorge Pérez-Juste, and Luis~M.
  Liz-Marzán.
\newblock Plasmonic {Supercrystals}.
\newblock {\em Acc. Chem. Res.}, 52(7):1855--1864, 2019.

\bibitem{BlancoFormoso2020}
Maria Blanco-Formoso, Nicolas Pazos-Perez, and Ramon~A. Alvarez-Puebla.
\newblock Fabrication of plasmonic supercrystals and their sers enhancing
  properties.
\newblock {\em ACS Omega}, 5:25485--25492, 2020.

\bibitem{Kreibig}
Uwe Michael~Vollmer Kreibig.
\newblock {(Springer series in materials science, v. 25) Uwe Kreibig{\_}
  Michael Vollmer-Optical properties of metal clusters-Springer (1995)}.

\bibitem{Kelly2003}
K.~Lance Kelly, Eduardo Coronado, Lin~Lin Zhao, and George~C. Schatz.
\newblock {The optical properties of metal nanoparticles: The influence of
  size, shape, and dielectric environment}.
\newblock {\em Journal of Physical Chemistry B}, 107(3):668--677, 2003.

\bibitem{MaierBook}
S.A. Maier.
\newblock {\em Plasmonics: Fundamentals and Applications}.
\newblock Springer, New York, 2007.

\bibitem{Meinzer2014}
Nina Meinzer, William~L. Barnes, and Ian~R. Hooper.
\newblock {Plasmonic meta-atoms and metasurfaces}.
\newblock {\em Nature Photonics}, 8(12):889--898, 2014.

\bibitem{Lamowski2018}
Simon Lamowski, Charlie~Ray Mann, Felicitas Hellbach, Eros Mariani, Guillaume
  Weick, and Fabian Pauly.
\newblock {Plasmon polaritons in cubic lattices of spherical metallic
  nanoparticles}.
\newblock {\em Physical Review B}, 97(12):1--11, 2018.

\bibitem{Barnes2003}
William~L. Barnes, Alain Dereux, and Thomas~W. Ebbesen.
\newblock {Surface plasmon subwavelength optics}.
\newblock {\em Nature}, 424(6950):824--830, 2003.

\bibitem{Shalaev2008}
Vladimir~M. Shalaev.
\newblock {Transforming Light}.
\newblock {\em Science}, 322(2):384--386, 2008.

\bibitem{Tame2013}
M.~S. Tame, K.~R. McEnery, Ş~K. {\"{O}}zdemir, J.~Lee, S.~A. Maier, and M.~S.
  Kim.
\newblock {Quantum plasmonics}.
\newblock {\em Nature Physics}, 9(6):329--340, 2013.

\bibitem{Baranov2020}
Denis~G. Baranov, Battulga Munkhbat, Elena Zhukova, Ankit Bisht, Adriana
  Canales, Benjamin Rousseaux, G{\"o}ran Johansson, Tomasz~J. Antosiewicz, and
  Timur Shegai.
\newblock Ultrastrong coupling between nanoparticle plasmons and cavity photons
  at ambient conditions.
\newblock {\em Nature Communications}, 11(1):2715, Jun 2020.

\bibitem{Kockum2019}
Anton~Frisk Kockum, Adam Miranowicz, Simone {De Liberato}, Salvatore Savasta,
  and Franco Nori.
\newblock {Ultrastrong coupling between light and matter}.
\newblock {\em Nature Reviews Physics}, 1(1):19--40, 2019.

\bibitem{FornDiaz2019}
P.~Forn-D\'{\i}az, L.~Lamata, E.~Rico, J.~Kono, and E.~Solano.
\newblock Ultrastrong coupling regimes of light-matter interaction.
\newblock {\em Rev. Mod. Phys.}, 91:025005, Jun 2019.

\bibitem{DeLiberato2014}
Simone {De Liberato}.
\newblock {Light-matter decoupling in the deep strong coupling regime: The
  breakdown of the purcell effect}.
\newblock {\em Physical Review Letters}, 112(1):1--5, 2014.

\bibitem{Artoni1991}
M~Artoni.
\newblock {Quantum-optical properties of polariton waves}.
\newblock {\em Physical Review B}, 44(8):3736, 1991.

\bibitem{Cirio2016}
Mauro Cirio, Simone {De Liberato}, Neill Lambert, and Franco Nori.
\newblock {Ground State Electroluminescence}.
\newblock {\em Physical Review Letters}, 116(11):1--7, 2016.

\bibitem{Hopfield1958}
J.~J. Hopfield.
\newblock {Theory of the contribution of excitons to the complex dielectric
  constant of crystals}.
\newblock {\em Physical Review}, 112(5):1555--1567, 1958.

\bibitem{Weick2015}
Guillaume Weick and Eros Mariani.
\newblock {Tunable plasmon polaritons in arrays of interacting metallic
  nanoparticles}.
\newblock {\em European Physical Journal B}, 88(1), 2015.

\bibitem{LeRuBook}
Eric C.~Le Ru and Pablo~G. Etchegoin.
\newblock {\em Principles of Surface-Enhanced Raman Spectroscopy}.
\newblock Elsevier, Amsterdam, 2009.

\bibitem{Reich2020}
Stephanie Reich, Niclas~S. Mueller, and Michal Bubula.
\newblock Selection rules for structured light in nanooligomers and other
  nanosystems.
\newblock {\em ACS Photon.}, 7:1537, 2020.

\bibitem{LiJensen2006}
Jensen Li, Gang Sun, and C.~T. Chan.
\newblock Optical properties of photonic crystals composed of metal-coated
  spheres.
\newblock {\em Phys. Rev. B}, 73:075117, Feb 2006.

\bibitem{Bohr1953}
Aage~Niels Bohr and Ben~R Mottelson.
\newblock {Collective and individual-particle aspects of nuclear structure},
  1953.

\bibitem{Bohr1998}
Aage Bohr and Ben~R. Mottelson.
\newblock {\em {Nuclear Structure, II - Buclear Deformations}}, volume~II.
\newblock World Scientific, Singapore, 1998.

\bibitem{Gulshani1978}
P~Gulshani.
\newblock {Exact Canonically Conjugate Momentum to the Quadrupole Tensor and
  Microscopic Derivation of the Nuclear Collective Hamiltonian}.
\newblock {\em Physics Letters}, 77(2):131--134, 1978.

\bibitem{Selsto2007}
S{\o}lve Selst{\o} and Morten F{\o}rre.
\newblock {Alternative descriptions of the light-matter interaction beyond the
  dipole approximation}.
\newblock {\em Physical Review A - Atomic, Molecular, and Optical Physics},
  76(2):4--7, 2007.

\bibitem{Kittel2005}
Charles Kittel.
\newblock {\em {Introduction to Solid State Physics}}.
\newblock Joh Wiley {\&} Sons, 2005.

\bibitem{Cohen1955}
M.~H. Cohen and F.~Keffer.
\newblock {Dipolar Sums in the Primitive Cubic Lattices}.
\newblock {\em Physical Review}, 99(4):1128--1134, 1955.

\bibitem{Hepp1973}
Klaus Hepp and Elliott~H Lieb.
\newblock On the superradiant phase transition for molecules in a quantized
  radiation field: the dicke maser model.
\newblock {\em Annals of Physics}, 76(2):360--404, 1973.

\bibitem{Savasta2020}
Salvatore Savasta, Omar {Di Stefano}, and Franco Nori.
\newblock {Thomas-Reiche-Kuhn (TRK) sum rule for interacting photons}.
\newblock {\em Nanophotonics}, 10(1):465--476, 2020.

\bibitem{Xiao2009}
Ming-wen Xiao.
\newblock {Theory of transformation for the diagonalization of quadratic
  Hamiltonians}.
\newblock {\em http://arxiv.org/abs/0908.0787}, 2009.

\bibitem{Kolwas2010}
K.~Kolwas.
\newblock {Plasmonic abilities of gold and silver spherical nanoantennas in
  terms of size dependent multipolar resonance frequencies and}.
\newblock {\em Opto-Electronics Review}, 22(2):77--85, 2010.

\bibitem{Shopa2010}
M.~Shopa, K.~Kolwas, A.~Dercachova, and G.~Dercachov.
\newblock {Dipole and quadrupole surface plasmon resonance contributions in
  formation of near-field images of a gold nanosphere}.
\newblock {\em Opto-Electronics Review}, 22(2):77--85, 2010.

\bibitem{Doerr2017}
T.~P. Doerr, O.~I. Obolensky, and Yi~Kuo Yu.
\newblock {Extending electrostatics of dielectric spheres to arbitrary charge
  distributions with applications to biosystems}.
\newblock {\em Physical Review E}, 96(6), 2017.

\bibitem{SunLin2018}
Lin Sun, Haixin Lin, Kevin~L. Kohlstedt, George~C. Schatz, and Chad~A. Mirkin.
\newblock Design principles for photonic crystals based on plasmonic
  nanoparticle superlattices.
\newblock {\em PNAS}, 115(28):7242--7247, 2018.

\bibitem{HuangChengPing2010}
Cheng-Ping Huang, Xiao-Gang Yin, Qian-Jin Wang, Huang Huang, and Yong-Yuan Zhu.
\newblock Long-wavelength optical properties of a plasmonic crystal.
\newblock {\em Phys. Rev. Lett.}, 104:016402, Jan 2010.

\bibitem{Purcell1946}
E.~M. Purcell, H.~C. Torrey, and R.~V. Pound.
\newblock Resonance absorption by nuclear magnetic moments in a solid.
\newblock {\em Phys. Rev.}, 69:37--38, Jan 1946.

\bibitem{Ciuti2005}
Cristiano Ciuti, G\'erald Bastard, and Iacopo Carusotto.
\newblock Quantum vacuum properties of the intersubband cavity polariton field.
\newblock {\em Phys. Rev. B}, 72:115303, Sep 2005.

\end{thebibliography}

\end{document}